# Social Complex Contagion in Music Listenership: A Natural Experiment with 1.3 Million Participants

**John Ternovski[1,2,3] and Taha Yasseri[3,4]**

[1] Yale University, New Haven, CT, USA
[2] Harvard Kennedy School, Harvard University, Cambridge, MA, USA
[3] Oxford Internet Institute, University of Oxford, Oxford, UK
[4] Alan Turing Institute, London, UK

Keywords: contagion, influence, social networks, indirect effects

**Abstract**:

Can live music events generate complex contagion in music streaming? This paper finds evidence in the affirmative—but only for the most popular artists. We generate a novel dataset from Last.fm, a music tracking website, to analyse the listenership history of 1.3 million users over a two-month time horizon. We use daily play counts along with event attendance data to run a regression discontinuity analysis in order to show the causal impact of concert attendance on music listenership among attendees and their friends network. First, we show that attending a music artist's live concert increases that artist's listenership among the attendees of the concert by approximately 1 song per day per attendee (p-value<0.001). Moreover, we show that this effect is contagious and can spread to users who did not attend the event. However, the extent of contagion depends on the type of artist. We only observe contagious increases in listenership for well-established, popular artists (.06 more daily plays per friend of an attendee [p<0.001]), while the effect is absent for emerging stars. We also show that the contagion effect size increases monotonically with the number of friends who have attended the live event.

**Significance statement**:

The internet has transformed the music industry's profit model, shifting the industry focus from record sales to live events and online streaming. However, to this date, there have been no large-scale, rigorous studies that quantify the causal impact of concerts on music listenership. This is the first natural experiment that discovers that concert attendance not only increases music listenership by attendees, but also spills over to the attendees' friends-networks. In some cases, the contagion effect appears to be larger in magnitude than the direct impact of attendance. In other words, if enough of your friends attend a concert, the impact on your music-listening behaviour may be larger than if you had attended the event yourself.

# Introduction

David Bowie has been quoted as saying "[m]usic itself is going to become like running water or electricity, [so] you'd better be prepared for doing a lot of touring" (*1*). Bowie's prescient prediction of a post-Napster world led one economist to coin the "Bowie Theory" to explain the changing economic model used by the music industry (*1*). Namely, the Bowie Theory summarizes the industry's shift from relying on revenue streams from physical copies of pre-recorded music to that of live performances. And recent research has found that while recorded sales have fallen, revenues from live performances have held steady (*2*, *3*). But does such an approach only work for the most popular bands? After all, music insiders claim that even "mid-level" bands would "be doing well to break even" (*4*). Altruistic motivations aside, why tour? Are there *any* economic benefits to live performances? Research on the music market has found that besides the direct benefits of touring (i.e. ticket and merchandise sales), tours provide an opportunity for an artist "to expand their fan base" (*5*). This latter "indirect effect" has generally been rather amorphous in the literature, but advances in network analysis allows us to frame this question in terms of contagion. In other words, do live events influence users who did not attend the event themselves, but are embedded in a social environment that experience "infection" (i.e. attendance of the live event by some members of the network)?

Social contagion has been studied in different systems and under different dispersion scenarios; these include political mobilization through peer networks (*6*, *7*), adoption of health behaviours among members of online communities (*8*) and real-world social networks (*9*), leveraging peers for viral marketing (*10*), and the spread of hash tags on Twitter (*11*, *12*). Social influence has also been found to have a critical role in the art market (*13*). But are the social contagion processes more effective for a certain category of artists? Social influence signals are widely used in such settings and help promote popular products to maximize market efficiency. However, it has been argued that social influence makes these markets unpredictable (*14*). As a result, social influence has been presented in a negative light. However, when it comes to market activities, such processes can lead to considerable revenues. The music industry in just the US has been valued at $17.2 billion as of 2016 (*15*), so there has been no shortage of incentives to optimize the revenue streams within it. However, it is the advent of illegal sharing of pre-recorded music by means of mp3s in the early 2000s that inspired renewed interest in developing new economic models for music consumption (*16*). In the modern music industry, there have been three major streams of revenue identified: (1) the Internet and digital music consumption, (2) CDs, records and conventional pre-recorded music consumption, and (3) live music consumption

(*17*). The new economic models that emerged gave greater weight to live performances (*18*) and the industry reacted. As of 2016, live music revenues accounted for more than half of all US music revenue (*15*). Prior to the 2000s, there have been relatively few studies that sought to understand the mechanics of live events (*5*). This is not surprising as the conventional wisdom in the music industry before the Internet was that live performances should be treated as nothing more than promotions for pre-recorded music (*3*).

Scholars identified the major reasons why artists choose to perform live as some combination of the following: (1) direct profits, (2) expanding listenership, and (3) strengthening their existing fan base (*5*). Prior to the Internet, it seems that the industry thought live performances could only satisfy the latter two factors, but, today, studies have clearly demonstrated that live music can capture a large share of direct revenues. Black et al. (*5*) find a trend of increasing ticket prices and unflagging demand, while industry reports illustrate live event revenues have continued to grow (*19*).

The question that remains is whether the traditional reason for concerts—namely, "promotional" effects—still exist. From a theoretical perspective, some researchers argue that there is something unique about live performances, as attendance has not waned despite the substantially higher costs (*20*). This may imply that despite major changes in the music industry and the increased convenience of digital music, live events are not simply a venue to sample music but a means by which individuals gain a new (or renewed) enthusiasm for the music (*20*). In other words, if the indirect promotional effects ever existed, it is likely that they still exist. Indeed, the increased ability to communicate information by the average concert-goer to a large number of personal contacts through online social networks may mean that the indirect effects are larger.

These indirect effects have been assumed by theoretical economic models (*18*), but actually connecting music consumption to offline behaviour has proven challenging due to sparse data availability. Montoro-Pons & Cuadrado-Garcia (*3*) made significant in-roads evaluating the link between concert attendance and music consumption, concluding that concert attendance does not cause increases in CD purchases. They do note that their analysis does not capture other modes of music consumption and concert attendance may still have impacts on music listenership (*3*). Another study linked pre-recorded music listenership with offline concert attendance to evaluate if fan preferences concord with songs actually played live (*21*). Most recently, Maasø (*22*) looked at streaming patterns before and after a music festival in Norway. However, that study was a macro-level analysis and so it did not attempt isolate the peer effects from other possible confounds or attempt a more stringent causal identification strategy (such as the regression discontinuity design we use in this paper).

We build upon the methodology used by Rodriguez et al. (*21*), which made use of Last.fm data to link listenership habits to live event attendance. Specifically, we evaluate if live events have any impact on music consumption. We also look at differences in impact by type of band (popular or "hyped") and differences in attendee demographics. More crucially, we investigate whether live event attendance has any *indirect* effects on listenership among the attendee's friends. We divide non-pecuniary benefits into either direct effects or indirect effects. Direct effects increase a given individual's music consumption of that band *as a result of* their attendance. These effects include the expansion of the band's fan base (i.e., non-fans who go to a live event and become fans, consuming the artist's music after the event) and satisfying existing fans (i.e., fans who see the live event and subsequently consume the artist's music at a greater rate). Indirect effects include all gains in music consumption from the live event by individuals *who did not themselves attend the event*. These effects capture the recommendations from event attendees and other forms of communication about the events.

We use Last.fm's song-listen data to capture an omnibus measure of music consumption and we rely on Last.fm self-reported live event attendance to capture our intervention of interest. As there are conflicting reports as to the profitability of touring depending on the artist's popularity, we also make the distinction between "popular" and "indie" ("Hyped") artists. To evaluate whether concerts actually have direct effects on the artist's fan base, we examine a given fan's listenership habits before and after live event attendance for 61,053 unique Last.fm users. We adopt a regression discontinuity design to determine if there are any shocks in listenership after event attendance, which would indicate that the performance either "re-activated" existing fans or generated new fans.

To evaluate indirect effects, we extract all of the attendees' Last.fm friends* (for a total of ~1.3 million users) and their listenership data. Any discontinuity in listenership at the time of the live event (that their friend(s) went to) can be interpreted as the indirect impact of event attendance on non-attendees. This measure captures indirect effects that result from either passive signalling or attendees actively recommending an artist to their friends.

We found strong evidence of direct effects for both Hyped and Top Artists. Attending an event of a Top Artist resulted in an immediate 1.13 song increase (p<0.001) in listenership by attendees. Similarly, Hyped Artist events were responsible for a 1.05 song increase (p<0.001) in their attendees' subsequent song listens. The indirect impacts were much more nuanced. Event attendance of Top Artists' concerts impacted the music listening behaviour of attendees' friends who did not attend the event by .06 songs (p<.001). While this is a very modest impact, we must keep in mind that the average concert-goer in our dataset has an average 56.8 friends just on the Last.fm network. And in some cases, more than one individual in a given social community will attend the same event. To determine if the indirect effects increase as the number of attendees increase, we varied the number of individuals who attended the event and examined the indirect effect on non-attending friends. We found evidence that the indirect effects increase as the number of concert-going friends increase. A user who had 5 friends, saw an indirect effect of 1.6 songs (p<.025), an effect ~150% greater the direct effect of attending a concert. We did not see any evidence of indirect effects for friends of attendees of Hyped Artist events.

---

* Last.fm has since adopted a follower/following social connection schema. At the time this study was conducted, Last.fm necessitated that both parties confirm their friendship to create a social link.

## Results

Following the Last.fm ontology, we bifurcate music artists into two categories: Hyped and Top Artists.[†] Hyped Artists have the largest *increases* in listenership, while Top Artists have the highest play counts (*23*). An example of a Top Artist in our dataset is Vampire Weekend, while the artist Yo-Yo Ma appears in our Hyped Artists list. For the two lists of artists, we extract live events between 2013 and 2014 documented on the Last.fm website; for each live event, we extract the list of all the Last.fm users who reported attending that event. Then we extract the attendees' basic demographic information, as well as all the songs they listened to a month before and after that live event. We also capture their entire Last.fm friends list and the listenership records of each of those friends. (This sampling strategy is a variant of "labelled star sampling" (*24*).)

To determine if there are differences in the characteristics of Hyped and Top Artist event attendees, we examine the summary statistics of the two sets of users. Identifying such differences helps build the case that music consumption and influence flows may be different across these two datasets. (Previous studies have found that fundamental demographic differences such as gender are associated with different motivations for attending live events *(25).)* As seen in Table 1, we find large differences in the gender composition of Hyped and Top Artist event attendees. Namely, Hyped Artists tend to attract ~54% of men to their shows, while Top Artists have a substantially more equitable gender breakdown with 49.7% of attendees reporting to be male. We should note that more than a tenth of our participants refrained from reporting their gender, so it is possible that it is not necessarily the actual gender composition of live events that differs across Top and Hyped bands, but rather the willingness to report a particular gender.

---

[†] For detailed definitions of these and other terms used, please see Material and Methods.

*Table 1: Demographic Statistics of Attendees*

| **Top Artist Dataset** | | | **Hyped Artist Dataset** | | |
|---|---|---|---|---|---|
| **Artists in Sample** | 85 | | **Artists in Sample** | 301 | |
| **Live Events** | 2,237 | | **Live Events** | 4,344 | |
| **Attendees** | 27,137 | | **Attendees** | 33,916 | |
| ***Attendees Self-Reported Gender*** | | | ***Attendees Self-Reported Gender*** | | |
| *Female* | 9,958 | *36.70%* | *Female* | 10,641 | *31.37%* |
| *Male* | 13,475 | *49.66%* | *Male* | 18,323 | *54.02%* |
| *Not Stated or Missing* | 3,704 | *13.65%* | *Not Stated or Missing* | 4,586 | *13.52%* |
| ***Attendees Self-Reported Location*** | | | ***Attendees Self-Reported Location*** | | |
| *United States* | 4,096 | *15.09%* | *United States* | 4,717 | *13.91%* |
| *United Kingdom* | 2,733 | *10.07%* | *United Kingdom* | 3,721 | *10.97%* |
| *Germany* | 1,749 | *6.45%* | *Germany* | 3,210 | *9.46%* |
| *Australia* | 1,465 | *5.40%* | *Poland* | 2,679 | *7.90%* |
| *Netherlands* | 1,021 | *3.76%* | *Netherlands* | 2,107 | *6.21%* |
| *Other, Not Stated, or Missing* | 16,073 | *59.23%* | *Other, Not Stated, or Missing* | 17,482 | *51.54%* |

We then look at the friends-network of these attendees. As seen in Table 2, the local metrics of the two networks are relatively similar, with similar numbers of friends on average. Our sampling strategy precludes us from being able to cite any statistics about the global network structure—particularly metrics of transitivity such as betweenness (*24*). Figure 1 illustrates one of the sub-networks that make up our data sample: a Metallica concert in Ecuador with all the self-reported Last.fm attendees (in blue) and their non-attending friends (in red).

*Table 2: Friends Network*

| | **Top Artist Dataset** | | **Hyped Artist Dataset** | |
|---|---|---|---|---|
| **Attendees & Friends of Attendees (node count)** | | 624,194 | **Attendees & Friends of Attendees (node count)** | 732,192 |
| **Friend Connections (edge count)** | | 1,493,536 | **Friend Connections (edge count)** | 1,714,049 |
| **Average Number of Friends (average degree)** | | 56.8 | **Average Number of Friends (average degree)** | 52.6 |

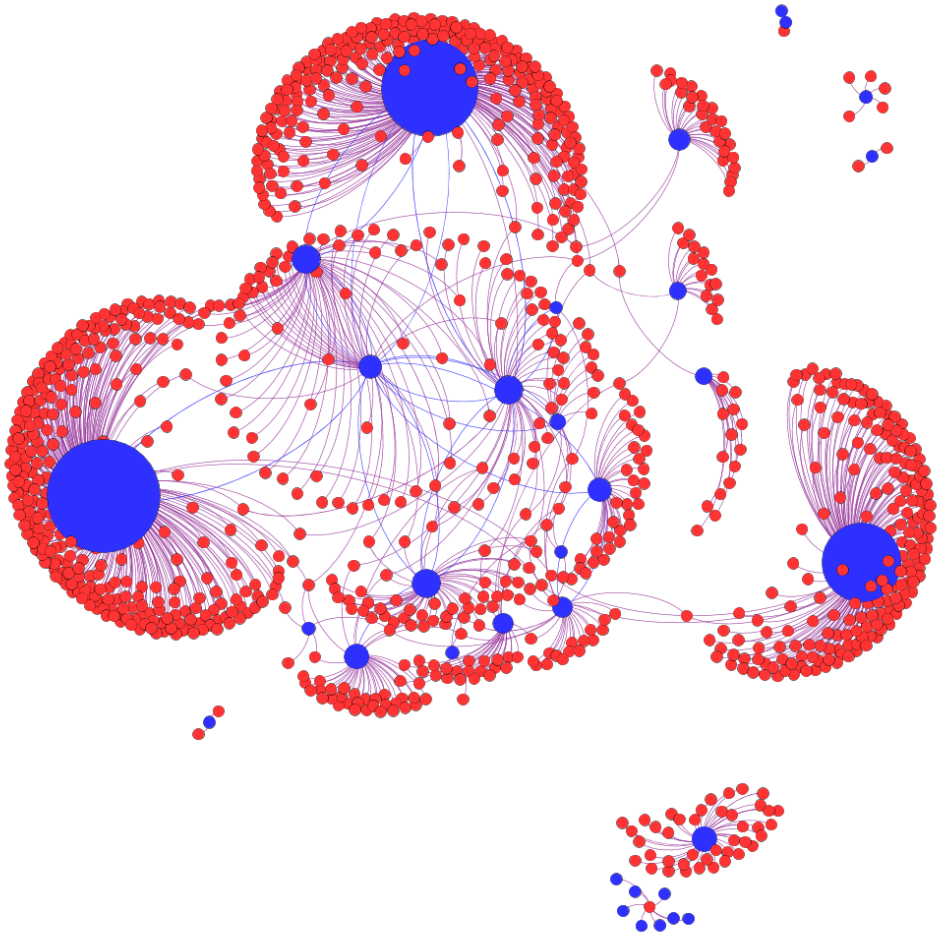

*Figure 1: Friends Network Visualization of Attendees of Metallica Concert in Quito, Ecuador on March 18, 2014*

**Direct Effects**

We first assess whether the live music event of a given artist increased listenership of the attendees of that artist's concert.‡ We find strong evidence of direct impacts on listenership among concert attendees of both Top and Hyped Artists. As seen in Table 3 and Figure 2, a Top Artist's live event increases listenership by 1.13 of a song (z-test p-value<.001), while a Hyped Artist live event increases listenership by 1.05 of a song (z-test p-value<.001).

---

‡ Among active users only; please see Materials and Methods for more details.

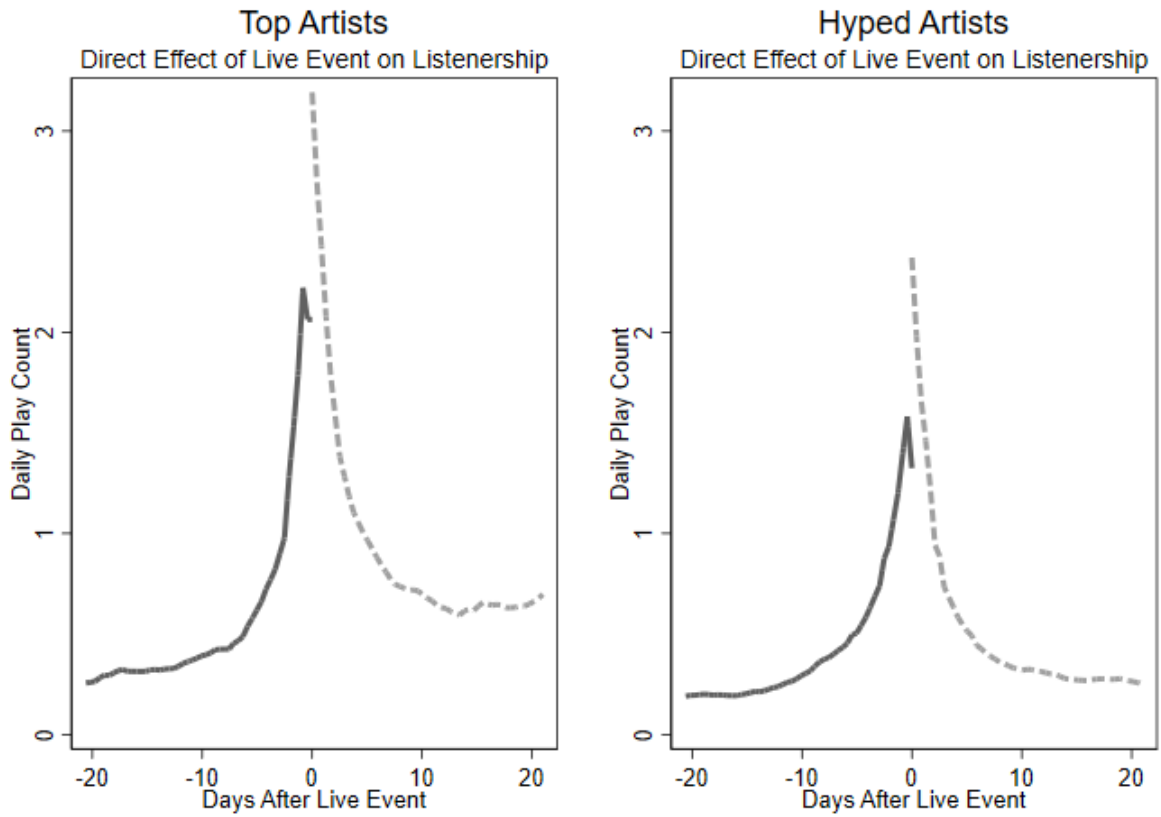

*Figure 2: Graphical Depiction of the Regression Discontinuity Estimate of the <u>Direct</u> Impact of Live Event*

*Table 3: Regression Discontinuity Estimate of the <u>Direct</u> Impact of Live Event*

|  | Top Artists | Hyped Artists |
|---|---|---|
| *impact of live event* | 1.13*** | 1.05*** |
| *s.e* | 0.08 | 0.08 |
| *Optimal Bandwidth* | 1.51 | 1.87 |

\*\*\* p-value<.001

Due to Last.fm's integration with Spotify, it is possible to estimate the impact of this increase in listenership on artist revenue.§ Since Spotify states that an artist earns on average $0.006 and $0.0084 per stream (*26*), if we assume each play is the average of these two figures, or $0.0072, we find that a Hyped Artist earns $0.0076 per average attendee in additional revenue from a live event. Similarly, a Top Artist earns an additional $0.0081 per average attendee. It is important to note that these impacts

---

§ Since the Last.fm song data includes not just Spotify song plays, but also users playing physical CDs and pirated material, the monetary estimates are used only to illustrate the magnitude of potential impacts.

are likely much larger. As seen in Figure 2, listenership increases before *and* after the event. For instance, within two days of Top Artist concerts, music listenership is on average 2.01 songs per day, while at all other points in our data, it is 0.55 songs per day; for Hyped Artists, it is 1.37 songs per day within two days of a concert and only 0.30 songs per day otherwise. However, we must emphasize that pre- and post-event increases in listenership are not causally identified.

As an example, one event included in the dataset of Top Artist events is a Taylor Swift concert at the O2 Arena in London, which had a reported attendance of 74,740 (*27*). This means that if 66%[**] of those attendees were Spotify listeners, this event generated an additional $401.34 from attendees' subsequent song listens.

**Indirect Effects**

Now that we have established that attending a live event impacts listenership habits of attendees, we want to determine if the attendees then, in turn, influence their friends, who have *not* attended the same event. Therefore, we apply the same regression discontinuity design to all friends of the attendees. One important note: to ensure we are looking at active users, we include only those users who have listened to the artist at least once in the 2-month observation window. As seen in Table 4 and Figure 3, we find a statistically significant impact on the listenership of the non-attending friends of attendees of Top Artists, but not Hyped Artists. The indirect effect on friends of Top Artist attendees is a .060 additional song plays, or more than 5% of the direct impact on listenership.[††] While this is a trivial impact on its own, it is important to emphasize that the mean Top Artist attendee has 56.8 friends.[‡‡] This means that one user's attendance translates to 3.4 more song plays on average, which translates to an increase of $0.025 per attendee. Using our earlier Taylor Swift concert example, the London concert secured an additional $1,210.40 in song streams. However, this analysis assumes only one friend attended the live event. We proceed by investigating whether the indirect impact increases as the number of friends who attended the event increases.

---

[**] We base this estimate on recent proprietary studies. A recent Spotify study found more than 2/3rds of attendees of a Dutch festival in 2014 used Spotify (*28*), while Music Watch Inc. found that 56% of Internet users streamed music in 2012 and 69% in 2014 (*29*).

[††] Non-attendees listen to the artist at an average of .47 songs per day within 2 days of the event and .41 at all other points in the dataset.

[‡‡] If some friends-lists are incomplete, this would bias our estimates of indirect economic effects downward.

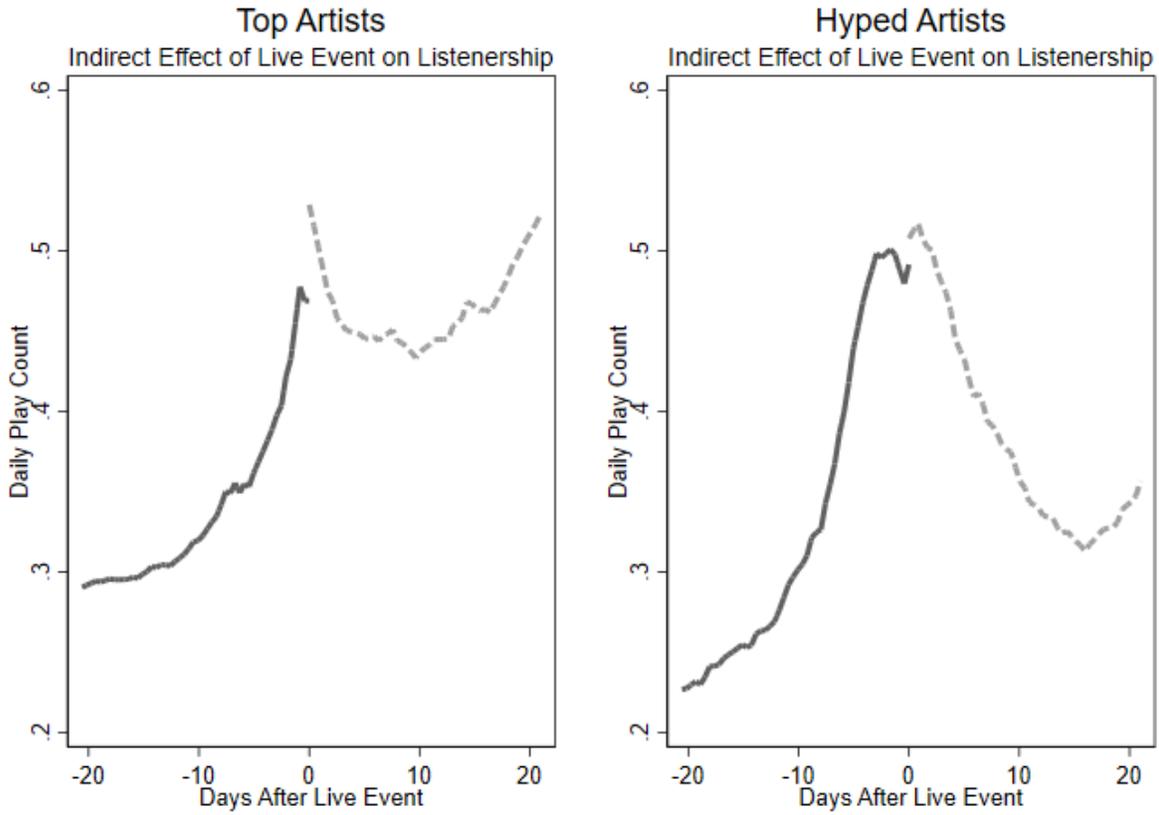

*Figure 3: Graphical Depiction of the Regression Discontinuity Estimate of the <u>Indirect</u> Impact of Live Event*

*Table 4: Regression Discontinuity Estimate of the <u>Indirect</u> Impact of Live Event*

|  | Top Artists | Hyped Artists |
|---|---|---|
| *impact of live event* | 0.060*** | 0.016 |
| *s.e.* | 0.012 | 0.025 |
| *Optimal Bandwidth* | 1.27 | 1.91 |

\*\*\* *p-value<.001*

Since the data includes non-attendees who are friends with multiple attendees, we can pursue a series of subgroup analyses. The fact that we have a partially observed network would normally preclude such an analysis, but we know the full list of Last.fm attendees, so we know that there is no unobserved Last.fm friend who actually attended the event. A key limitation of our analysis is that event attendance is self-reported. It is very possible that a given individual appears to be a non-attendee in our data, but simply failed to report their event attendance. However, there is no reason why this measurement error could not go in the opposite direction, as well. Non-attendees can just as easily either incorrectly report their attendance (either as a social desirability lie or to simply demonstrate their intention of going). Future studies should investigate the prevalence of these effects to identify the direction and magnitude of the measurement error.

To determine whether the number of attendee-friends interacts with the indirect effect, we run a series of regression discontinuity analyses across non-attending users with various numbers of attendee-friends. We find that as the number of friends who attended the event increases, the influence on listenership increases monotonically. To better discern if this pattern stems from multiple attendees exerting influence on the non-attender, we perform a permutation test. Namely, we look at the discontinuity estimates across the number of attendee-friends in our actual dataset and then compare these results to a synthetic dataset where we held the friend network constant but randomly assigned that non-attending friend to a different live event date of the same artist in the same 30-day period such that no user in their friend network attended that live event.[§§] Table 5 and Figure 4 illustrate these results.

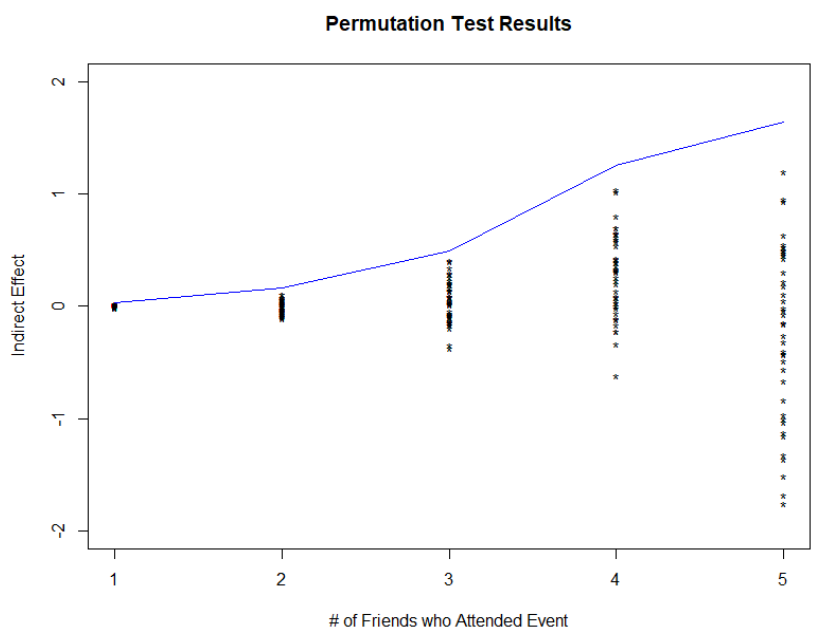

*Figure 4: Graphical Depiction of Permutation Test Results*

---

[§§] We use the same bandwidth generated for the main indirect effects analysis to make comparisons across subgroups. Our analysis is robust to using a leave-one-out cross validation approach separately for each subgroup, though the results are weaker.

*Table 5: Permutation Test Results*

| # Friends who Attended Event | Observed Effect | Null Distribution Average | Permutation Test p-value (one-sided) |
|---|---|---|---|
| 1 | 0.033 | 0.005 | <.025 |
| 2 | 0.164 | -0.008 | <.025 |
| 3 | 0.493 | 0.041 | <.025 |
| 4 | 1.258 | 0.276 | <.025 |
| 5 | 1.639 | -0.243 | <.025 |

*We exclude all instances of non-attendees with more than 5 friends due to the likelihood that the non-attending friend may have simply failed to report attendance. (See for instance the cluster of attendees in the lower right corner in Figure 2.) Less than 0.2% of our sample have more than 5 friends who attended the same event.*

Even with the permutation test, we are not able to ascertain if there is a *causal* relationship between the number of friends and the increasing indirect effect. It is quite possible that having multiple friends attend a given live event is indicative of the level of marketing/advertisements about the event (i.e. more marketing may induce more members of a given social circle to attend an event). However, the impact of marketing is unlikely to be tied to the day of the event. Rather, marketing about a concert is much more likely to occur in the days leading up to the event, which may explain the rapid increases in listenership immediately before our discontinuities.

**Subgroup Analysis**

As a means of exploring further hypotheses, we conduct subgroup analysis across attendee's individual-level characteristics. Since these hypotheses were not made *a priori,* due to the risk of data-dredging (*30*), we do not look at any statistical tests across groups. Rather, we hope to get a Bayesian baseline for potential differences, which should be evaluated rigorously in future studies.

Since prior studies have shown that the motivations of live event attendees tend to differ across standard demographic characteristics such as gender (*24*, *31*), we check whether the impact of live events on music consumption differs across location and gender.[***] Because there are clear differences in demographic characteristics between the attendees of Top and Hyped Artist shows, we pursue subgroup analysis separately for each batch of data. As seen in Table 6, we see few meaningful differences across attendee demographics. While, there are some slight differences in the direct impact of Top Artist live events on self-identified males as compared to self-identified females, we should not make any rigorous comparisons of these statistics, as there may be large biases in gender self-reporting. If this difference is true however, this would mean that males have greater direct impacts of concert attendance.

---

[***] While users can also share their age, this data was not available on Last.fm's API at the time this analysis was conducted.

*Table 6: Direct Impacts in Listenership Across Attendee Demographics Measured by Daily Song Plays*

| Hyped | | | | Top | | | |
|---|---|---|---|---|---|---|---|
| Male | Female | US | non-US | Male | Female | US | non-US |
| 0.47 | 0.58 | 0.91 | .96 | 1.26 | 0.92 | 1.11 | 1.07 |

We also investigate whether the demographics of non-attending users translate to differential indirect effects (Table 7). In this case, we find that there are larger indirect impacts on non-US, non-attending users for both Hyped and Top artists. There is a slightly larger indirect effect on female non-attending friends in the Hyped Artist universe, but it is important to note that the indirect effects in the Hyped Artist universe were not significant.

*Table 7: Indirect Impacts in Listenership across Non-attending Friend Demographics Measured by Daily Song Plays*

| Hyped | | | | Top | | | |
|---|---|---|---|---|---|---|---|
| Male | Female | US | non-US | Male | Female | US | non-US |
| 0.036 | 0.054 | -0.002 | 0.028 | 0.054 | 0.051 | 0.027 | 0.066 |

# Discussion and Conclusion

This paper illustrates that even without any profits from tickets and merchandise sales at shows, there may be long-term economic benefits for touring bands. As we have shown, there are sizable, statistically significant impacts on music consumption of attending a live event. The size of the *direct* impact is comparable for both Top Artists and Hyped artists. However, while there are substantial indirect effects for Top Artists, there are no statistically significant effects for the up-and-coming artists. In concordance with the Black et. al study (*5*), we see that the rich get richer, while the average Hyped Artist struggles to expand their fanbase through touring.

One remarkable finding is that, if enough of your friends see a live event, there is a bigger boost in listenership of that artist than even if you yourself attended the event. This suggests that the word-of-mouth effects have the potential to be more important in indirect revenues than perhaps even direct ticket sales. One potential confound that we must be cautious of is that it is very possible that the non-attending listener may have actually attended the event. After all, if very many of a given user's friends attended the event, it may be indicative that the users themselves attended the event. Future studies should use alternative measures of attendance to assess the overall accuracy of concert-attendance self-reports.

Additionally, it would be useful to better understand the means by which indirect effects spill over to non-attendees. Do attendees of live events actually talk to their friends about their experience at the concert or is the direct effect of listenership spilling over to non-attendee friends? In other words, the play counts of an attendee are public, so are non-attendee friends seeing the increased listenership and, in turn, listening to more of that artist themselves? Additionally, a user's attendance of a live event is also publicly available—are non-attendees seeing that their friend attended a live event by a given artist and so they listen to the artist themselves?

Finally, future studies should also analyse just how much listenership increases across all (monetized) streaming platforms. Our results only cover those individuals who use Last.fm and have enabled music tracking on their preferred streaming service. These individuals may be systematically different from the usual concert attendee/streaming-music-listener.

## Materials and Methods

The analysis in this paper uses publicly available data from the music website Last.fm. We use three distinct types of data: (1) track listens, (2) event attendance, and (3) the Last.fm friends network.

**Track Listens Data**

Last.fm is a free music website with over 20 million active users (*32*) that keeps track of the songs played by its members. A user does not have to listen to the music directly from the Last.fm website for it to be recorded (or "scrobbled") in the user's track history—a user needs only to install the Audioscrobbler plugin on their music software (e.g., Itunes, Windows Media Player, etc.). The plugin keeps track of the music, as well as the time and date when the track is played—even when the user is not connected to the Internet. When the user next reconnects to the Internet, the stored data is dumped onto Last.fm's servers with date information updating retroactively. We extract this data to be used as our main outcome variable of interest.

In 2014, Last.fm partnered with Spotify (*33*), which led to the inclusion of the Audioscrobbler plug-in in any Spotify installation by default, such that the user just needed to switch on Last.fm music scrobbling in their Spotify settings page.

The plugin and the API can be manipulated by users with programming experience to record tracks that someone has not actually listened to. We exclude cases where, in a given day, the individual appeared to have listened to a given artist for a duration that is greater than the length of a day. The average song length of contemporary popular music is 227 seconds (*34*) and so more than 380 songs in a day is unlikely to translate into actual listens.[†††]

For the purposes of this study, we made the a priori decision to look at a two-month time horizon—with one month of listenership data prior to the attendance of an event and one month of listenership data after the attendance of the event. Since we are unable to easily establish whether a Last.fm user is active, we apply our regression discontinuity only to users who listened to at least one song by the target artist in the 2-month observation window.

**Event Attendance Data**

Last.fm also serves as a platform for publicizing live events, which can be uploaded by fans, promoters, or the artists themselves. Last.fm listeners can indicate that they are going to attend or have attended an event. The event pages then retain the exact date and time of the event along with all the attendees. Events prior to the existence of Last.fm can also be uploaded (e.g. the band Deep Purple have events as far back as 1968 with 8 self-reported attendees). To avoid recall bias (i.e., where users will retroactively mark that they attended only the most memorable events), we concentrated on recent live events. In 2014, we extracted live events from "Top Artists" and "Hyped Artist" that occurred January, 2013 - October, 2014. Hyped Artists correspond to bands who have the highest rate of growth while Top Artists are the most listened to artists on the site overall (*23*). We decided that "hype" is the best, easily-available proxy for "indie" artists who are recently popular enough to yield a large enough sample size of attendees, but are still far from the "Top Artist" charts.

---

[†††] All analyses are robust to the inclusion of these outliers.

We aimed to get comparable sample sizes of attendees in the Top Artist and the Hyped Artists datasets, so we ended up extracting the live events of 85 Top Artists and 300 Hyped Artists. (Top Artists had more documented live events and higher levels of self-reported attendance.) We exclude artists that have not actually had live events in recent years (e.g. Bob Dylan). We also insure that no artists are found both in the Hyped and Top Artist datasets. We only include events that had one eligible Hyped or Top Artist playing.

**"Friends" Data**

Last.fm also serves as a music-based social network; users are able to "friend" other people and discuss music and artists primarily through "shoutboxes." Research on these friend networks indicated that users primarily friend other users with similar musical taste (*35*), which makes this network particularly susceptible to indirect effects.

We extracted a network of friends for all attendees of events by Hyped and Top artists. (We suspect there were scraping errors for some attendees resulting in the extraction of an incomplete friends list. All analyses are robust to excluding these suspect attendees.) We then extracted two months of listening history (one month before and after the date of the live event) of the corresponding artist for each friend of each attendee.

**Regression Discontinuity Design**

In order to assess the direct impact of live events on attendees' listenership, we use a regression discontinuity design, as is standard practice for similar phenomena (*36*, *37*).[‡‡‡] We report all results using the bandwidth generated using the leave-one-out cross validation approach to minimize squared bias and variance (*39*). As recommended in Imbens, & Lemieux (*36*), we use the rectangular kernel for all analyses and verify its robustness using a triangular kernel. All analyses with the triangular kernel exhibit similar results. Since the same Last.fm user could have attended multiple events by a Top Artist in our dataset, we cluster standard errors on attendee. The results do not change meaningfully when we do not cluster the standard errors. Because each of these regression discontinuities used a slightly different bandwidth, as a robustness check, we use the Top Artist bandwidth with the Hyped Artist regression discontinuity and vice versa; the reported discontinuity impacts do not change meaningfully.

---

[‡‡‡] We use the Stata package "rd" coded by Austin Nichols (*38*).

# Acknowledgments

TY was partially supported by The Alan Turing Institute under the EPSRC grant EP/N510129/1.

# References

1. Krueger, A. B. (2005). The economics of real superstars: The market for rock concerts in the material world. *Journal of Labor Economics*, 23(1), 1–30.
2. Charron, J. P. (2017). Music audiences 3.0: Concert-goers' psychological motivations at the dawn of virtual reality. *Frontiers in Psychology*, *8*.
3. Montoro-Pons, J. D., & Cuadrado-García, M. (2011). Live and prerecorded popular music consumption. *Journal of Cultural Economics*, *35*(1), 19-48.
4. Passman, D. S. (2006). *All You Need to Know About the Music Business* Free Press 6th ed.
5. Black, G. C., Fox, M. A., & Kochanowski, P. (2007). Concert tour success in North America: An examination of the top 100 tours from 1997 to 2005. *Popular Music and society*, *30*(2), 149-172.
6. Bond, R. M., Fariss, C. J., Jones, J. J., Kramer, A. D., Marlow, C., Settle, J. E., & Fowler, J. H. (2012). A 61-million-person experiment in social influence and political mobilization. *Nature*, *489*(7415), 295-298.
7. Jones JJ, Bond RM, Bakshy E, Eckles D, Fowler JH (2017) Social influence and political mobilization: Further evidence from a randomized experiment in the 2012 U.S. presidential election. *PLoS ONE,* 12(4): e0173851. https://doi.org/10.1371/journal.pone.0173851
8. Centola, D. (2010). The spread of behavior in an online social network experiment. *Science*, 329(5996), 1194-1197.
9. Aral, S., & Nicolaides, C. (2017). Exercise contagion in a global social network. *Nature Communications*, *8*, 14753. http://doi.org/10.1038/ncomms14753
10. Leskovec, J., Adamic, L. A., & Huberman, B. A. (2007). The dynamics of viral marketing. *ACM Transactions on the Web (TWEB)*, *1*(1), 5.
11. Romero, D. M., Meeder, B., & Kleinberg, J. (2011, March). Differences in the mechanics of information diffusion across topics: idioms, political hashtags, and complex contagion on twitter. In *Proceedings of the 20th international conference on World wide web* (pp. 695-704). ACM.
12. Mønsted B, Sapieżyński P, Ferrara E, Lehmann S (2017) Evidence of complex contagion of information in social media: An experiment using Twitter bots. *PLoS ONE,* 12(9): e0184148. https://doi.org/10.1371/journal.pone.0184148
13. Salganik, M. J., & Watts, D. J. (2008). Leading the herd astray: An experimental study of self-fulfilling prophecies in an artificial cultural market. *Social Psychology Quarterly*, *71*(4), 338-355.
14. Salganik, M. J., Dodds, P. S., & Watts, D. J. (2006). Experimental study of inequality and unpredictability in an artificial cultural market. *Science*, 311(5762), 854-856.
15. Resnikoff, Paul. (2017, June 7) "Live Concerts + Streaming = 73% of the US Music Industry" *Digital Music News* Retrieved from https://www.digitalmusicnews.com/2017/06/07/music-industry-concerts-streaming/
16. Liebowitz, Stan (2003), "Will MP3 Downloads Annihilate the Record Industry? The Evidence so Far," in Gary Libecap (ed.), *Advances in the Study of Entrepreneurship, Innovation, and Economic Growth*, JAI Press.
17. Rogers, J. (2013). The Death and Life of the Music Industry in the Digital Age. London: Bloomsbury Academic.
18. Dewenter, R., Haucap, J., & Wenzel, T. (2012). On file sharing with indirect network effects between concert ticket sales and music recordings. *Journal of Media Economics*, *25*(3), 168-178.
19. Washenko, A. (2017, June 8). PwC Entertainment and Media Outlook, Part 2: United States results. *RAIN News*. Retrieved from http://rainnews.com/pwc-entertainment-and-media-outlook-part-2-united-states-results/

20. Earl, P. E. (2001). Simon's travel theorem and the demand for live music. *Journal of Economic Psychology*, 22 (3), 335–358.
21. Rodriguez, M. A., Gintautas, V., & Pepe, A. (2008). A Grateful Dead analysis: The relationship between concert and listening behavior. *First Monday* 14 (1)
22. Maasø, A. (2016). Music Streaming, Festivals, and the Eventization of Music. *Popular Music and Society*, 1-22.
23. Whiting, D. (2012, December 18). So this is Christmas, and what have we done? *Last.fm*
24. Kolaczyk, E. (2009). *Statistical Analysis of Network Data*. Springer.
25. Bowen, H. E., & Daniels, M. J. (2005). Does the music matter? Motivations for attending a music festival. *Event Management*, *9*(3), 155-164.
26. Spotify. (n.d.). How is Spotify contributing to the music business? Retrieved from http://www.spotifyartists.com/spotify-explained/#how-we-pay-royalties-overview
27. The Red Tour, (n.d). In *Wikipedia.* Retrieved Spetember 27, 2015, from https://en.wikipedia.org/wiki/The_Red_Tour
28. Page, W. (2014). Adventures in the Lowlands: Spotify, Social Media and Music Festivals. *Spotify Insights.* Retrieved from https://insights.spotify.com/us/2014/10/22/adventures-in-the-lowlands/
29. Crupnick, R. (2015). From Stream to Ticket: Mapping the Value of Music Discovery. Music Watch Inc. Retrieved from http://www.musicwatchinc.com/blog/from-stream-to-ticket-mapping-the-value-of-music-discovery/
30. Assmann, S. F., Pocock, S. J., Enos, L. E., & Kasten, L. E. (2000). Subgroup analysis and other (mis) uses of baseline data in clinical trials. *The Lancet*, 355(9209), 1064-1069. Retrieved from http://www.sciencedirect.com/science/article/pii/S0140673600020390http://www.sciencedirect.com/science/article/pii/S0140673600020390
31. Pegg, S., & Patterson, I. (2010). Rethinking music festivals as a staged event: Gaining insights from understanding visitor motivations and the experiences they seek. In *Journal of Convention & Event Tourism* (Vol. 11, No. 2, pp. 85-99). Taylor & Francis Group.
32. Weber, H. (2012, December 6). Spotify announces 5M+ paid subscribers globally, 1M paid in US, 20M total active users, 1B playlists. *TNW.* Retrieved from https://thenextweb.com/insider/2012/12/06/spotify-announces/
33. Dredge, S. (2014, January 30). Last.fm plots streaming music comeback with Spotify and YouTube. *The Guardian.* Retrieved from http://www.theguardian.com/technology/2014/jan/30/lastfm-streaming-comeback-spotify-youtube
34. Anisko, N. & Anderson, E. (2012). Average Length of Top 100 Songs on iTunes. *StatCrunch*. Retrieved from http://www.statcrunch.com/5.0/viewreport.php?reportid=28647&groupid=948
35. Baym, N. K., & Ledbetter, A. (2009). Tunes that bind? Predicting friendship strength in a music-based social network. *Information, Communication & Society*, *12*(3), 408-427.
36. Imbens, G. W., & Lemieux, T. (2008). Regression discontinuity designs: A guide to practice. *Journal of Econometrics*, *142*(2), 615-635.
37. Jacob, R. T., Zhu, P., Somers, M. A., & Bloom, H. S. (2012). *A Practical Guide to Regression Discontinuity*. MDRC.
38. Nichols, Austin. 2011. rd 2.0: Revised Stata module for regression discontinuity estimation. Retrieved from http://ideas.repec.org/c/boc/bocode/s456888.html
39. Imbens, G., & Kalyanaraman, K. (2012). Optimal bandwidth choice for the regression discontinuity estimator. *The Review of Economic Studies*, *79*(3), 933-959.